\documentclass{ws-procs975x65}           
\usepackage{amsmath,amsfonts,amssymb}
\usepackage{bm}

\newcommand{\dE}{I \! \! E}
\newcommand{\dB}{I \! \! B}

\newcommand{\dF}{I \! \! F}

\begin{document}

\title{Spin in gravitational and electromagnetic fields}

\author{Yuri N. Obukhov}

\address{Russian Academy of Sciences, Nuclear Safety Institute (IBRAE),\\
B. Tulskaya 52, 115191 Moscow, Russia\\
E-mail: obukhov@ibrae.ac.ru}

\begin{abstract}
A unified approach to the study of classical and quantum spin in external fields is developed. Understanding the dynamics of particles with spin and dipole moments in arbitrary gravitational, inertial and electromagnetic fields is important in astrophysics and high-energy and heavy-ion experimental physics. 
\end{abstract}


\bodymatter

\section{Projection method for spin in arbitrary external fields}\label{pro}

One can develop a natural extension of the classic Frenkel-Thomas spin model which is based on the definition of the ``magnetic'' and ``electric'' components of the relativistic forces and moments of forces acting on a particle (hence it is natural to call this approach a projection method). Let us recall that in Maxwell's theory, electrodynamic phenomena are described by the field strength tensor $F_{\alpha\beta}$, and the magnetic and electric fields can be introduced in a covariant way as the longitudinal and transversal projections on the velocity $U^\alpha$:
\begin{equation}
E_\alpha := F_{\alpha\beta}U^\beta,\qquad B^\alpha := {\frac 1{2c}}\eta^{\alpha\mu\nu\beta}
F_{\mu\nu}U_\beta.\label{EBmax}
\end{equation}
Here $\eta_{\alpha\beta\mu\nu}$ is the totally antisymmetric Levi-Civita tensor. Thereby one gets an unique representation of the electromagnetic field strength tensor in terms of its projections (electric and magnetic fields):
\begin{equation}
F_{\alpha\beta} = {\frac 1{c^2}}\left(E_\alpha U_\beta - E_\beta U_\alpha + c\eta_{\alpha\beta\mu\nu}
U^\mu B^\nu\right).\label{FEB}
\end{equation}

In the model of a classical particle with internal degrees of freedom (the generalized Frenkel-Thomas model), the motion of a test particle is characterized by the 4-velocity $U^\alpha$ and the 4-vector of spin $S^\alpha$, which satisfy the normalization $U_{\alpha} U^\alpha = c^2$ and the orthogonality condition $S_{\alpha} U^\alpha = 0$. In general, the dynamic equations for these variables can be written as
\begin{eqnarray}
{\frac {dU^\alpha}{d\tau}} = {\cal F}^\alpha,\qquad
{\frac {dS^\alpha}{d\tau}} = \Phi^\alpha{}_\beta S^\beta.\label{dotUS}
\end{eqnarray}
External fields of different physical nature (electromagnetic, gravitational, scalar, etc.) determine the forces ${\cal F}^\alpha$ acting on the particle, as well as the spin transport matrix $\Phi^\alpha{}_\beta$ which affects the spin. Normalization and orthogonality of the velocity and spin vectors impose conditions on the right-hand sides of (\ref{dotUS}):
\begin{equation}
U_\alpha {\mathcal F}^\alpha = 0,\qquad U_\alpha\Phi^\alpha{}_\beta S^\beta =
-\,S_\alpha {\mathcal F}^\alpha.\label{cc}
\end{equation}
The spin transport matrix must be skew-symmetric, $\Phi_{\alpha\beta} = - \Phi_{\beta\alpha}$, which automatically guarantees $S_\alpha S^\alpha=$const. The relativistic 4-velocity vector $U^\alpha$ is conveniently parametrized via the spatial velocity 3-vector $\widehat{v}^a$ and the Lorentz factor $\gamma$:
\begin{equation}\label{U}
U^\alpha = \left(\begin{array}{c}\gamma \\ \gamma \widehat{v}^a\end{array}\right),\qquad
\gamma = {\frac {1}{\sqrt{1 - \widehat{v}^2/c^2}}},\qquad \widehat{v}^2 = \delta_{ab}\widehat{v}^a\widehat{v}^b.
\end{equation} 

The vector of the generalized force ${\mathcal F}^\alpha$ acting on a particle has only a transversal projection, according to (\ref{cc}). As for the spin transport matrix $\Phi_{\alpha\beta}$, like any bivector, we can decompose it into a pair of 4-vectors. Namely, by analogy with (\ref{EBmax}), we define ``electric'' and ``magnetic'' projections
\begin{equation}
N_\alpha := \Phi_{\alpha\beta}U^\beta,\qquad Q^\alpha := {\frac 1{2c}}\eta^{\alpha\mu\nu\beta}
\Phi_{\mu\nu}U_\beta.\label{NQ}
\end{equation}
By construction, these 4-vectors are orthogonal to the velocity of the particle,
\begin{equation}
U_\alpha N^\alpha = 0,\qquad U_\alpha Q^\alpha = 0,\label{cc1}
\end{equation}
and the relations (\ref{cc}) are then recast into
\begin{equation}
U_\alpha {\mathcal F}^\alpha = 0,\qquad S_\alpha N^\alpha = S_\alpha {\mathcal F}^\alpha.\label{cc2}
\end{equation}
As a result, we have an unique decomposition analogous to (\ref{FEB}):
\begin{equation}
\Phi_{\alpha\beta} = {\frac 1{c^2}}\left(N_\alpha U_\beta - N_\beta U_\alpha + c\eta_{\alpha\beta\mu\nu}
U^\mu Q^\nu\right).\label{PNQ}
\end{equation}
The orthogonality conditions (\ref{cc1}) can be resolved so that
\begin{equation}
N^0 = {\frac {1}{c^2}}(\bm{N}\cdot\widehat{\bm{v}}),\label{N0}
\end{equation}
where as usual $\widehat{\bm{v}} = \{\widehat{v}^a\}$ and $\bm{N} = \{N^a\}$. Accordingly, we find 
in components:
\begin{eqnarray}
\Phi^a{}_0 &=& \gamma\left(N^a - \widehat{v}^aN^0 - \epsilon^a{}_{bc}Q^b\widehat{v}^c\right),\label{P0a}\\
\Phi^0{}_a &=& {\frac {\gamma}{c^2}}\left(N_a - \widehat{v}_aN^0 - \epsilon_{abc}Q^b\widehat{v}^c\right),\label{Pa0}\\
\Phi^a{}_b &=& {\frac {\gamma}{c^2}}\left(\widehat{v}^aN_b - \widehat{v}_bN^a\right)
- \gamma\epsilon^a{}_{bc}\left(Q^c - \widehat{v}^cQ^0\right).\label{Pab}
\end{eqnarray}

The physical spin, as an ``internal angular momentum'' of a particle, is defined with respect to particle's rest frame system, in which $u^\alpha = \delta^\alpha_0$. The transition to this system is carried out by the Lorentz transformation $U^\alpha = \Lambda^\alpha{}_\beta u^\beta$, where
\begin{equation}
\Lambda^\alpha{}_\beta = \left(\begin{array}{c|c}\gamma & \gamma\widehat{v}{}_b/c^2 \\
\hline \gamma\widehat{v}{}^a & \delta^a_b + {\frac {\gamma - 1}
{\widehat{v}{}^2}}\,\widehat{v}{}^a\widehat{v}{}_b\end{array}\right).\label{Lam}
\end{equation}
Therefore, the dynamic equation for the {\it physical spin} $s^\alpha = (\Lambda{}^{- 1})^\alpha{}_\beta S^\beta$ reads:
\begin{equation}
{\frac {ds^\alpha}{d\tau}} = \Omega^\alpha{}_\beta s^\beta,\label{dsdt}
\end{equation}
where the tensor of angular precession of spin is constructed as
\begin{equation}\label{Omab1}
\Omega^\alpha{}_\beta = (\Lambda^{-1})^\alpha{}_\gamma\Phi^\gamma{}_\delta\Lambda^\delta{}_\beta 
- (\Lambda^{-1})^\alpha{}_\gamma{\frac d {d\tau}} \Lambda^\gamma{}_\beta.
\end{equation}
The 0th component of (\ref{dsdt}) vanishes, which is equivalent to the second condition (\ref{cc}). As a result, the spin evolution equation (\ref{dsdt}) reduces to the 3-vector form
\begin{equation}
{\frac {ds^a}{d\tau}} = \Omega^a{}_b s^b,\qquad \hbox{or}\qquad
{\frac {d{\bm s}}{d\tau}} = {\bm \Omega}\times{\bm s}.\label{ds1}
\end{equation}
Here the components of 3-vectors are introduced through ${\bm s} = \{s^a\}$ and ${\bm\Omega} = \left\{-\,{\frac 12}\epsilon^{abc}\Omega_{bc}\right\}$. A direct calculation gives the precession angular velocity in terms of the magnetic and electric projections of the spin transport matrix:
\begin{equation}
\bm{\Omega} = \gamma\,\Bigl\{\bm{Q} - \widehat{\bm v}\,Q^0 + {\frac \gamma{\gamma + 1}}
\,{\frac {1}{c^2}}\left[\widehat{\bm v}\,(\widehat{\bm v}\cdot\bm{Q})  - \widehat{v}{}^2\bm{Q}
\right]\Bigr\} + {\frac {\gamma}{\gamma + 1}}\,{\frac {\bm{\mathcal F}\times\widehat{\bm v}}{c^2}}.\label{OG0}
\end{equation}
It is worthwhile to note that $\bm{\Omega}$ actually depends only on the transversal part of the ``magnetic'' vector, namely $ Q_\bot^\alpha = Q^\alpha - {\frac 1 {c^2}} U^\alpha U_\beta Q^\beta $. Explicitly in components
\begin{equation}
Q_\bot^0 = {\frac {\gamma^2}{c^2}}\left[\widehat{\bm v}\cdot\bm{Q} - \widehat{v}{}^2Q^0\right],\qquad
Q_\bot^a = Q^a + {\frac {\gamma^2}{c^2}}\left[\widehat{\bm v}\cdot\bm{Q} - c^2Q^0\right]
\widehat{v}^a,\label{QQ}
\end{equation}
and by a direct computation one can check that
\begin{equation}
\bm{Q}_\bot - \widehat{\bm v}\,Q_\bot^0 = \bm{Q} - \widehat{\bm v}\,Q^0,\qquad
\widehat{\bm v}\,(\widehat{\bm v}\cdot\bm{Q}_\bot)  - \widehat{v}{}^2\bm{Q}_\bot = 
\widehat{\bm v}\,(\widehat{\bm v}\cdot\bm{Q})  - \widehat{v}{}^2\bm{Q}.\label{vQ}
\end{equation}
From (\ref{QQ}) one then finds
\begin{equation}
Q_\bot^0 = {\frac 1{c^2}}\,\widehat{\bm v}\cdot\bm{Q}_\bot,\label{Qbot} 
\end{equation}
and therefore we recast (\ref{OG0}) into a final form 
\begin{equation}
\bm{\Omega} = \bm{Q}_\bot - {\frac {\gamma}{\gamma + 1}}\,{\frac {\widehat{\bm v}\,(\widehat{\bm v}\cdot\bm{Q}_\bot)}{c^2}} - {\frac {\gamma}{\gamma + 1}}\,{\frac {\widehat{\bm v}\times\bm{\mathcal F}}{c^2}}.\label{OG}
\end{equation}
The new general equations (\ref{OG}), (\ref{ds1}) are valid for a particle with spin that interacts with arbitrary external fields. The actual dynamics of the physical spin depends on the forces acting on the particle and on the spin transport law.

\section{Application: Spin in the gravitational and electromagnetic fields}

In order to consider the most general case, we assume that the gravitational field is described by an arbitrary coframe and an (independent) local Lorentz connection $(e^\alpha_i, \Gamma_i{}^{\alpha\beta} = -\Gamma_i{}^{\beta\alpha})$, whereas the electromagnetic field as usual is represented by the vector potential $A_i = (-\,\Phi, \bm{A})$. The corresponding field strengths are the torsion $T_{ij}{}^\alpha = \partial_ie^\alpha_j - \partial_je^\alpha_i + \Gamma_{i\beta}{}^\alpha e^\beta_j - \Gamma_{j\beta}{}^\alpha e^\beta_i$, the curvature $R_{ij}{}^{\alpha\beta} = \partial_i\Gamma_j{}^{\alpha\beta}- \partial_j\Gamma_i{}^{\alpha\beta} + \Gamma_{i\gamma}{}^\beta\Gamma_j{}^{\alpha\gamma} - \Gamma_{j\gamma}{}^\beta\Gamma_i{}^{\alpha\gamma}$, and the Maxwell tensor $F_{ij} = \partial_iA_j - \partial_jA_i$. Accordingly, the spacetime geometry in general carries the Riemann-Cartan structure with the metric $g_{ij} = e^\alpha_i e^\beta_j g_{\alpha\beta} $ and the nontrivial torsion. The ``deviation'' of the spacetime geometry from the Riemannian structure can be conveniently measured by the contortion tensor which is defined in terms of the difference of the local Lorentz connection $\Gamma_i{}^{\alpha\beta}$ and the Riemannian (Levi-Civita or Christoffel) connection $\tilde{\Gamma}_i{}^{\alpha\beta}$:
\begin{equation}
K_i{}^{\alpha\beta} = \tilde{\Gamma}_i{}^{\alpha\beta} - \Gamma_i{}^{\alpha\beta}.\label{GGK}
\end{equation}

\subsection{Quantum spinning particle}

A fermion particle with spin ${\frac 12}$, charge $q$ and mass $m$ is described by the relativistic Dirac theory. The spinor field $\psi$ dynamics is determined by the Lagrangian
\begin{eqnarray}\label{LD}
L = {\frac {i\hbar}{2}}\left(\overline{\psi}\gamma^\alpha D_\alpha\psi
- D_\alpha\overline{\psi}\gamma^\alpha\psi\right) - mc \overline{\psi}\psi 
+ {\frac {1}{2c}}M_{\alpha\beta}\,\overline{\psi}\sigma^{\alpha\beta}\psi
+ {\frac {i\hbar\nu'}{12}}\check{T}^\alpha\,\overline{\psi}\gamma_\alpha\gamma_5\psi,
\end{eqnarray}
where the spinor covariant derivative is defined as (with $\sigma_{\alpha\beta} = i\gamma_{[\alpha} \gamma_{\beta]}$)
\begin{equation}
D_\alpha\psi = e_\alpha^i \left(\partial _i\psi - {\frac {iq}{\hbar}}\,A_i\psi
+ {\frac i4}\Gamma_i{}^{\beta\gamma}\sigma_{\beta\gamma}\psi\right).\label{eqin2}
\end{equation}
The first two terms in (\ref{LD}) describe minimal coupling of the spinor field to electromagnetism and gravity encoded in (\ref{eqin2}). In addition, we assume possible non-minimal interactions, which are described by two Pauli-type terms in (\ref{LD}), where
\begin{equation}\label{Mab}
M_{\alpha\beta} = \mu' F_{\alpha\beta} + c\delta'\,{\frac 12}\,\eta_{\alpha\beta\mu\nu}F^{\mu\nu},
\end{equation}
is the generalized polarization tensor, and the axial torsion is defined as
\begin{equation}
\check{T}{}^\alpha = -\,{\frac 12}\,\eta^{\alpha\mu\nu\beta}T_{\mu\nu\beta} = \left\{\check{T}{}^0, \bm{\check{T}}\right\}.\label{axitor}
\end{equation}
The parameters $\delta', \mu', \nu'$ characterize the strength of nonminimal couplings.

\subsection{Classical spinning particle}

The classical theory of spin was developed soon after the concept of spin was proposed in particle physics (see\cite{corben} for introduction and history). This theory underlies the analysis of the dynamics of polarized particles in accelerators and storage rings.

Neglecting second-order spin effects\cite{chicone}, the dynamical equations for a spinning particle in external electromagnetic and gravitational fields are written as
\begin{eqnarray}
{\frac {DU^\alpha}{d\tau}} &=& -\,{\frac qm}\,F^\alpha{}_\beta\,U^\beta,\label{eomU}\\
{\frac {DS^\alpha}{d\tau}} &=& -\,(\nu'\! - 2)\,U^iK_{i\beta}{}^\alpha S^\beta 
- {\frac qm}\,F^\alpha{}_\beta S^\beta \nonumber\\
&& -\,{\frac 2\hbar}\left[M^\alpha{}_\beta + {\frac {1}{c^2}}U^\gamma\left(
U^\alpha M_{\beta\gamma}  - U_\beta M^\alpha{}_\gamma\right)\right]S^\beta\!.\label{eomS}
\end{eqnarray}
Here we follow the notations and conventions introduced in the previous sections.

\subsection{Application of the projection method}

Comparing the system (\ref{eomU})-(\ref{eomS}) with (\ref{dotUS}), we find explicitly the generalized force and the spin transport matrix:
\begin{eqnarray}
{\mathcal F}_\alpha = -\,{\frac {q}{m}}\,\dF_{\alpha\beta}\,U^\beta,\qquad
\Phi_{\alpha\beta} = -\,{\frac {q}{m}}\,\dF_{\alpha\beta},\label{FePe}
\end{eqnarray}
where we introduced the combined external field
\begin{equation}
\dF_{\alpha\beta} = F_{\alpha\beta} + {\frac {2m}{q\hbar}}\,M_{\alpha\beta}^\bot
- {\frac {m}{q}}\,U^i\Gamma_{i\alpha\beta}.\label{Fc}
\end{equation}
As we see, $\Phi_{\alpha\beta}U^\beta = {\mathcal F}_\alpha$. Here we denoted 
\begin{equation}
M_{\alpha\beta}^\bot = \Bigl(\delta_\alpha^\mu - {\frac {1}{c^2}}U_\alpha U^\mu\Bigr)
\Bigl(\delta_\beta^\nu - {\frac {1}{c^2}}U_\beta U^\nu\Bigr)M_{\mu\nu}.\label{Mbot} 
\end{equation}
The components of this tensor read explicitly
\begin{eqnarray}
M_{0a}^\bot &=& {\frac {\gamma^2}{c}}\left\{-\,v^2{\mathcal P}_a + \widehat{v}_a(\widehat{\bm v}
\cdot\bm{\mathcal P}) + c[\widehat{\bm v}\times\bm{\mathcal M}]_a\right\},\label{Mb0}\\
M_{ab}^\bot &=& \epsilon_{abc}\,{\frac {\gamma^2}{c^2}}\left\{c^2{\mathcal M}^c - \widehat{v}^c
(\widehat{\bm v}\cdot\bm{\mathcal M}) + c[\widehat{\bm v}\times\bm{\mathcal P}]^c\right\}.\label{Mba}
\end{eqnarray}
Here we identify $c\bm{\mathcal P}{}_a = \{M_{\widehat{0}\widehat{1}}, M_{\widehat{0}\widehat{2}}, M_{\widehat{0}\widehat{3}} \}, \bm{\mathcal M}{}^a = \{ M_{\widehat{2}\widehat{3}}, M_{\widehat{3}\widehat{1}}, M_{\widehat{1}\widehat{2}} \}$ as the polarization and magnetization 3-vectors, respectively. 

Now we use the projection method developed above and extract the ``electric'' and ``magnetic'' vectors from the artificial electromagnetic field (\ref{Fc})
\begin{equation}
\dE_\alpha := \dF_{\alpha\beta}U^\beta,\qquad \dB^\alpha := {\frac 1{2c}}
\eta^{\alpha\mu\nu\beta}\dF_{\mu\nu}U_\beta,\label{EBc}
\end{equation}
which yields an unique representation
\begin{equation}
\dF_{\alpha\beta} = {\frac 1{c^2}}\left(\dE_\alpha U_\beta - \dE_\beta U_\alpha
+ c\eta_{\alpha\beta\mu\nu}U^\mu\dB^\nu\right).\label{FEB1}
\end{equation}
Comparing (\ref{NQ}) with (\ref{PNQ}), we thus identify 
\begin{eqnarray}
N_\alpha = -\,{\frac qm}\,\dE_\alpha,\qquad
Q^\alpha = -\,{\frac qm}\,\dB^\alpha.\label{NEQB}  
\end{eqnarray}

Let us define effective ``electric'' and ``magnetic'' fields
\begin{equation}\label{EBeff}
\bm{\mathfrak{E}}{}_a^{\rm eff} = \{ \dF_{\widehat{1}\widehat{0}}, \dF_{\widehat{2}\widehat{0}},
\dF_{\widehat{3}\widehat{0}} \},\qquad\bm{\mathfrak{B}}{}^a_{\rm eff} = \{\dF_{\widehat{2}\widehat{3}},
\dF_{\widehat{3}\widehat{1}}, \dF_{\widehat{1}\widehat{2}} \}.
\end{equation}
Or in compact form: $\bm{\mathfrak{E}}{}_a^{\rm eff} = \dF_{a0}$, and $\bm{\mathfrak{B}}{}^a_{\rm eff} = {\frac 12}\epsilon^{abc}\dF_{bc}$. 

Consequently from (\ref{EBc}) and (\ref{NEQB}) we derive the components 
\begin{eqnarray}
N^0 = \gamma\,{\frac {q\widehat{\bm v}\cdot\bm{\mathfrak{E}}{}_{\rm eff}}{mc^2}}, &\qquad&
\bm{N} = \gamma\,{\frac qm}\left(\bm{\mathfrak{E}}{}_{\rm eff} + \widehat{\bm v}\times
\bm{\mathfrak{B}}{}_{\rm eff}\right),\label{Neff}\\
Q^0 = -\,\gamma\,{\frac {q\widehat{\bm v}\cdot\bm{\mathfrak{B}}{}_{\rm eff}}{mc^2}}, &\qquad&
\bm{Q} = -\,\gamma\,{\frac qm}\Bigl(\bm{\mathfrak{B}}{}_{\rm eff} - {\frac 1{c^2}}
\widehat{\bm v}\times\bm{\mathfrak{E}}{}_{\rm eff}\Bigr).\label{Qeff}
\end{eqnarray}
By construction, $\bm{Q}_\bot = \bm{Q}$, and moreover we have $\bm{\mathcal F} = \bm{N}$. Inserting (\ref{Neff}) and (\ref{Qeff}) into (\ref{OG}), we obtain the precession angular velocity (\ref{OG})
\begin{equation}\label{Otot}
\bm{\Omega} = {\frac qm}\left( -\,\bm{\mathfrak{B}}{}_{\rm eff} + {\frac {\gamma}
{\gamma + 1}}\,{\frac {\widehat{\bm{v}}\times\bm{\mathfrak{E}}{}_{\rm eff}}{c^2}}\right)
\end{equation}
as a function of external fields which enter via the effective variables: 
\begin{eqnarray}
\bm{\mathfrak{E}}_{\rm eff} &=& \bm{\mathfrak{E}} - {\frac {2m}{q\hbar}}\,\gamma^2\,
\widehat{\bm{v}}\times\bm{\Delta} + {\frac mq}\,\bm{\mathcal E},\label{EEE}\\
\bm{\mathfrak{B}}_{\rm eff} &=& \bm{\mathfrak{B}} + {\frac {2m}{q\hbar}}\,\gamma^2
\Bigl[\bm{\Delta} - {\frac 1{c^2}}\,\widehat{\bm{v}}\,(\widehat{\bm{v}}\cdot\bm{\Delta})
\Bigr] + {\frac mq}\,\bm{\mathcal B}.\label{BBB}
\end{eqnarray}
Recall that the components of the true electric and magnetic fields are introduced as $\bm{\mathfrak{E}}{}_a = \{F_{\widehat{1}\widehat{0}}, F_{\widehat{2}\widehat{0}}, F_{\widehat{3}\widehat{0}} \}, \bm{\mathfrak{B}}{}^a = \{F_{\widehat{2}\widehat{3}}, F_{\widehat{3}\widehat{1}}, F_{\widehat{1}\widehat{2}} \}$. The generalized polarization current,
\begin{equation}\label{Delta}
\bm{\Delta} = \bm{\mathcal M} + {\frac 1c}\,\widehat{\bm{v}}\times\bm{\mathcal P},
\end{equation}
accounts in the spin precession (\ref{Otot})-(\ref{BBB}) for the electromagnetic nonminimal coupling effects, whereas the gravitoelectric and gravitomagnetic fields
\begin{equation}
\bm{\mathcal E} = \widetilde{\bm{\mathcal E}} - {\frac {\gamma c(3 - \nu')}{6}}\,\widehat{\bm{v}}
\times\check{\bm{T}},\qquad \bm{\mathcal B} = \widetilde{\bm{\mathcal B}} + {\frac {\gamma c
(3 - \nu')}{6}}\left(\check{\bm{T}} - \widehat{\bm{v}}\check{T}{}^0\right)\label{EBTN}
\end{equation}
encompass general-relativistic Riemannian gravitoelectric $\widetilde{\bm{\mathcal E}}$ and gravitomagnetic $\widetilde{\bm{\mathcal B}}$ contributions, as well as the post-Riemannian terms due to the spacetime torsion.

\section{Discussion and conclusion}

The classical and quantum spin dynamics are fully consistent. The physical contents of the relativistic quantum theory (\ref{LD}) is revealed when we recast the Dirac equation into a Schr\"odinger form and go to the Foldy-Wouthuysen (FW) representation. In the Schwinger gauge, the coframe is parametrized by the functions $V, \bm{K}, W^{\widehat a}{}_b$:
\begin{equation}\label{coframe}
e_i^{\,\widehat{0}} = V\,\delta^{\,0}_i,\qquad e_i^{\widehat{a}} =
W^{\widehat a}{}_b\left(\delta^b_i - cK^b\,\delta^{\,0}_i\right),\qquad a=1,2,3,
\end{equation}
and the resulting FW Hamiltonian in the semiclassical approximation then reads
\begin{equation}
{\mathcal H}_{FW} = \beta mc^2V\gamma + q\Phi + {\frac c2}\left(\bm{K}\cdot\bm{\pi}
+ \bm{\pi}\cdot\bm{K}\right) + {\frac \hbar 2}\bm{\Sigma}\cdot\bm{\Omega}.\label{Hamlt}
\end{equation}
Here the Lorentz factor (\ref{U}) operator $\gamma$ and the precession velocity (\ref{Otot}) operator $\bm{\Omega}$ are both expressed in terms of the velocity operator $\widehat{\bm v}$ which is related to the momentum operator $\bm\pi=-i\hbar\bm{\nabla} - q\bm A$ via $\beta\,W^b{}_{\widehat a}\pi_b = m\gamma\widehat{v}{}_a$.

The analysis of the spin dynamics in electromagnetic, inertial and gravitational fields is fundamentally important for the study of the geometrical structure of spacetime\cite{ost4}, as well as in the high-energy physics experiments\cite{ost5}, in the search of gravitational waves\cite{ost6}, in the neutrino physics in matter\cite{max}, and in the heavy-ion collisions\cite{pro}. 

This work was partially supported by the Russian Foundation for Basic Research (Grant No. 18-02-40056-mega).


\begin{thebibliography}{99}

\bibitem{corben}
H.C. Corben, {\it Classical and quantum theories of spinning particles}
(Holden-Day, Inc: San Francisco, 1968).

\bibitem{chicone}
C. Chicone, B. Mashhoon, B. Punsly,
{\sl Phys. Lett. A} {\bf 343}, 1 (2005).

\bibitem{ost4}
Yu.N. Obukhov, A.J. Silenko, O.V. Teryaev,
{\sl Phys. Rev. D} {\bf 90}, 124068 (2014).

\bibitem{ost5}
Yu.N. Obukhov, A.J. Silenko, O.V. Teryaev,
{\sl Phys. Rev. D} {\bf 94}, 044019 (2016).

\bibitem{ost6}
Yu.N. Obukhov, A.J. Silenko, O.V. Teryaev,
{\sl Phys. Rev. D} {\bf 96}, 105005 (2017).

\bibitem{max}
M. Dvornikov, 
{\sl Phys. Rev. D} {\bf 99}, 035027 (2019).

\bibitem{pro}
G.Y. Prokhorov, O.V. Teryaev, V.I. Zakharov,
{\sl Phys. Rev. D} {\bf 98}, 071901(R) (2018).

\end{thebibliography}
\end{document}